\documentclass[referee,usegraphicx]{mn2e}

\def\beq{\begin{eqnarray}}
\def\eeq{\end{eqnarray}}
\def\e{{\rm e}}
\def\d{{\rm d}}
\def\M{{\cal M}}
\def\MM{{\cal M}^2}
\def\Msh{{\cal M}_{\rm sh}}
\def\Mmin{{\cal M}_{\rm min}}
\def\v{\upsilon}
\def\vr{\upsilon_r}
\def\vphi{\upsilon_\varphi}
\def\om{\omega}
\def\omp{\omega'}
\def\p{\partial}
\def\cs{c_{\rm s}}

\def\dvr{\delta\upsilon_r}
\def\dvphi{\delta\upsilon_\varphi}
\def\dcs{\delta c_{\rm s}}
\def\drho{\delta\rho}
\def\rsh{r_{\rm sh}}
\def\rs{r_{\rm son}}
\def\rco{r_{\rm co}}
\def\fsh{f_{\rm sh}}
\def\gsh{g_{\rm sh}}
\def\dSsh{\delta S_{\rm sh}}
\def\dS{\delta S}
\def\Bsh{B_{\rm sh}}

\def\Tac{\tau_{\rm ac}}
\def\Tadv{\tau_{\rm adv}}
\def\na{\nabla}
\def\Pa{P_{\rm an}}
\def\Pl{P_{\rm lin}}
\def\Ps{P_{\rm sim}}

\title[Non-axisymmetric instabilities in shocked accretion flows]
{Non-axisymmetric instabilities in shocked adiabatic accretion flows}

\author[Wei-Min Gu and Ju-Fu Lu]
{Wei-Min Gu \thanks{E-mail: guwm@xmu.edu.cn} and Ju-Fu Lu
\\
Department of Physics, Xiamen University, Xiamen 361005, China
}

\begin{document}

\maketitle

\begin{abstract}
We investigate the linear stability of a shocked accretion flow
onto a black hole in the adiabatic limit.
Our linear analyses and numerical calculations show that,
despite the post-shock deceleration, the shock is generally unstable
to non-axisymmetric perturbations.
The simulation results of Molteni, T\'oth \& Kuznetsov can be
well explained by our linear eigenmodes.
The mechanism of this instability is confirmed to be based on the
cycle of acoustic waves between the corotation radius and the shock.
We obtain an analytical formula to calculate the oscillation period
from the physical parameters of the flow.
We argue that the quasi-periodic oscillation should be a common
phenomenon in accretion flows with angular momentum.
\end{abstract}

\begin{keywords}
accretion, accretion discs -- black hole physics -- hydrodynamics
-- instabilities -- shock waves
\end{keywords}

\section{Introduction}

Hydrodynamic instabilities of shocked accretion flows may explain
quasi-periodic oscillation (QPO) processes occurring in black hole
candidates.
The structure of stationary black hole accretion flows involving
standing shocks was first described by Fukue (1987). Subsequently,
shock studies have been made extensively in both inviscid and viscous
accretion flows (Chakrabarti \& Das 2004; Gu \& Lu 2004 and
references therein).
However, even with the simple inviscid hypothesis,
the stability of the shock is not fully understood.
In the isothermal limit, Nakayama (1992) introduced a global
instability between a sonic point and a shock, and
found the criterion that ``post-shock acceleration causes instability",
which was confirmed by the simulations of Nobuta \& Hanawa (1994).
Moreover, Nakayama (1994) investigated this instability in an
adiabatic flow and claimed that such a criterion is also correct unless
the shock is extremely strong.
All the above works, however, were only for axisymmetric perturbations.

The pioneering work of Papaloizou \& Pringle (1984) found a
non-axisymmetric instability based on the acoustic cycle between
the corotation radius and the boundary.
The Papaloizou-Pringle instability (hereafter PPI) is
known to take place in discs or tori, in which the radial velocity
is initially zero.
The mechanism of such an instability has been discussed extensively by
Goldreich \& Narayan (1985), Goldreich, Goodman \& Narayan (1986),
Narayan, Goldreich \& Goodman (1987) and Kato (1987).
The effect of radial advection on the PPI was investigated
by Blaes (1987), who found that the PPI is strongly stabilized by
advection at the inner boundary.
Another type of non-axisymmetric instability was found in a spherical
accretion flow by Foglizzo \& Tagger (2000) and Foglizzo (2001, 2002),
which is based on the cycle of entropy/vorticity perturbations and
acoustic waves in the subsonic region between a
stationary shock and a sonic surface.
Recently, Gu \& Foglizzo (2003, hereafter Paper I) studied a shocked
accretion disc in the isothermal limit and found that
the shock is generally linearly unstable to non-axisymmetric
perturbations despite the post-shock deceleration.
Paper I pointed out that such an instability is a form of PPI
modified by advection and the presence of the shock.
Apart from the above linear works, Molteni, T\'oth \& Kuznetsov
(1999, hereafter MTK) performed 2-D simulations of a shocked adiabatic
flow and found a non-axisymmetric instability.
MTK simulations showed that the instability saturates at a low level,
and a new asymmetric configuration develops, with a deformed shock
rotating steadily.
The mechanism of the instability, however, was not explained
in MTK. They briefly mentioned a possible link with the
non-axisymmetric disc instabilities studied by Blaes \& Hawley (1988).

In this paper, we investigate the stability of a shocked accretion
flow in the adiabatic and inviscid limit.
We compare our linear results with
the non-linear simulation results of MTK and manage to explain
the different behaviours in their simulations.
The paper is organized as follows.
In Section 2, we present a set of linearized equations and boundary
conditions. Subsequently, in section 3, we present the linear
results by numerically solving the equations. Finally, in section 4,
we summarize our conclusions and present a discussion.

\section{Equations}

An adiabatic flow around a black hole is considered
in the pseudo-Newtonian potential introduced by Paczy\'nski and
Wiita (1980), $\Phi \equiv -GM/(r-r_{\rm g})$.
Equations are made dimensionless by using the Schwarzschild radius
and the speed of light as reference units, i.e., $r_{\rm g} \equiv 1$
and $c \equiv 1$.
In this paper, the thickness of the flow is approximated as a constant
for the sake of simplicity, as in Nakayama (1994), Blaes (1987),
MTK and Paper I.

The stationary flow is described by the conservation of mass and the
Bernoulli equation:
\beq
&& \rho r \vr = {\rm const.} \ ,
\label{mass} \\
&& e = \frac{\vr^2}{2} + \frac{l^2}{2r^2} + \frac{\cs^2}{\gamma-1}
- \frac{1}{2(r-1)} = {\rm const.} \ ,
\label{bernoulli}
\eeq
where $\rho$ is the density, $\vr$ is the radial velocity,
$\cs = (\gamma p/\rho)^{1/2}$ is the sound speed,
$l$ is the specific angular momentum, and
$e$ is the Bernoulli constant.
The structure of a stationary flow involving a standing shock can be
obtained from the above equations by a given pair of $(l,e)$
(see Appendix A of MTK for details).

The continuity equation and the Euler equation are
written as follows:
\beq
&& \frac{\p\rho}{\p t} + \na \cdot (\rho\v) = 0 \ ,
\label{masscon} \\
&& \frac{\p\v}{\p t} + w\times \v + \na \left[ \frac{\v ^2}{2}
+ \frac{\cs^2}{\gamma-1} - \frac{1}{2(r-1)} \right]
= \cs ^2 \na \frac{S}{\gamma} \ ,
\label{Euler}
\eeq
where $w$ is the vorticity, and $S$ is the entropy.

Apart from vorticity perturbations and acoustic waves in
an isothermal flow, entropy perturbations should appear in
an adiabatic flow. Thus the linearized equations here are
a little more complicated than those in Paper I.
In order to write out the linearized equations in the simplest form,
the two functions $f,g$ are defined as follows:
\beq
&& f \equiv \vr\dvr + \frac{2}{\gamma-1}\cs\dcs + \vphi\dvphi \ ,
\label{feq} \\
&& g \equiv \frac{\drho}{\rho} + \frac{\dvr}{\vr} \ ,
\label{geq}
\eeq
where $f$ is the perturbation of the Bernoulli constant and $g$ is the
perturbation of the mass accretion rate.
The frequency $\omp$ measured in the rotating frame is defined as:
\beq
&& \omp \equiv \om - m\Omega \ ,
\eeq
where $\om$ is the complex frequency of the perturbation,
$m$ is the azimuthal wave number,
and $\Omega \equiv l/r^2$ is the angular velocity.
With the standard method of linear stability analysis, i.e.,
assuming perturbations to be proportional to $e^{-i\om t+im\varphi}$,
the following differential system is obtained:
\beq
\frac{\p f}{\p r} = && -\frac{i\omp\MM}{\vr(1-\MM)}f
+ \frac{i\om\vr}{1-\MM}g   \nonumber \\
&& + i\om\vr\left(\frac{1}{1-\MM}+\frac{1}{\gamma\MM}\right)
\dSsh \e^{\int_{\rsh}^r\frac{i\omp}{\vr}\d r}   \nonumber \\
&& + \frac{l}{r^2\vr(1-\MM)}
\Bsh \e^{\int_{\rsh}^r\frac{i\omp}{\vr}\d r} \ ,
\label{dfdr} \\
\frac{\p g}{\p r} = && \frac{i}{\om\vr}
\left[\frac{\omp^2}{\cs^2(1-\MM)}-\frac{m^2}{r^2}\right]f
-\frac{i\omp\MM}{\vr(1-\MM)}g   \nonumber \\
&& -\frac{i\omp}{\vr(1-\MM)}
\dSsh \e^{\int_{\rsh}^r\frac{i\omp}{\vr}\d r}   \nonumber \\
&& -\frac{1}{\om r^2\vr}\left[m+\frac{\omp l}{\cs^2(1-\MM)}\right]
\Bsh \e^{\int_{\rsh}^r\frac{i\omp}{\vr}\d r} \ ,
\label{dgdr}
\eeq
where $B \equiv r\vr w_z - {im\cs ^2\dS / \gamma }$, $w_z$ is
the vorticity along the rotation axis,
$\M \equiv -\vr /\cs$ is the radial Mach number, and the
subscript ``sh" denotes the shock position.
The boundary conditions corresponding to a perturbed shock velocity
$\Delta v_r$ are obtained:
\beq
&& \fsh = (\v_+ - \v_-) \Delta \vr \ ,
\label{fsh} \\
&& \gsh = \frac{\omp}{\om}\left(\frac{1}{v_+}-\frac{1}{v_-}\right)
\Delta \vr \ ,
\label{gsh} \\
&& \dSsh = -\frac{\gamma(\v_+ - \v_-)^2}{c_{\rm s+}^2\v_-}
\left[ \frac{\omp}{\om} + \frac {i\v_+}{\om}\frac{\d \ln \M_+}{\d r}
\right] \Delta \vr \ ,
\label{dSsh} \\
&& \Bsh = 0 \ ,
\label{Bsh}
\eeq
where the subscripts ``-" and ``+" denote the pre-shock and
post-shock values, respectively. \\
In addition to the two boundary conditions
Eqs.(\ref{fsh}-\ref{gsh})
at the shock, a third equation is obtained from the critical
condition at the sonic point,
\beq
&& -\frac {\omp}{\om \vr^2} f_{\rm son} + g_{\rm son}
+ \dSsh \e^{\int_{\rsh}^{\rs}\frac{i\omp}{\vr}\d r} = 0 \ ,
\label{critic}
\eeq
where the subscript ``son" denotes the sonic point.
These three boundary conditions are used to numerically solve the
differential system Eqs.~(\ref{dfdr}-\ref{dgdr}) and to determine the
eigenfrequencies $\om$.

The methods for obtaining the above linearized equations and
boundary conditions are similar to those in Paper I
(see Appendices B and C of Paper I for details).

\section{Numerical results}

\begin{table*}
\begin{center}
\caption[]{A sample of 15 shocked accretion flows
for numerical calculations.}
\label{case15}
\begin{tabular}{crccccc}
\hline
case & $\rsh$ & $l$ & $\Msh$ & $\Mmin$ & $m=1$ linear results
& MTK simulation results \\
\hline
1  &  5.3 & 1.7770 & 0.651 & 0.650 & unstable & regular oscillation \\
2  &  5.3 & 1.8000 & 0.574 & 0.572 & unstable & regular oscillation \\
3  &  5.3 & 1.8100 & 0.539 & 0.536 & unstable & beating \\
4  &  5.3 & 1.8225 & 0.493 & 0.489 & unstable & irregular \\
5  &  7.8 & 1.8000 & 0.629 & 0.556 & unstable & regular oscillation \\
6  &  7.8 & 1.8100 & 0.595 & 0.520 & unstable & beating \\
7  &  7.8 & 1.8200 & 0.560 & 0.482 & unstable & beating \\
8  & 12.7 & 1.8200 & 0.684 & 0.465 & unstable & regular oscillation \\
9  & 17.2 & 1.8255 & 0.760 & 0.439 & unstable & nearly stable \\
10 & 23.4 & 1.8620 & 0.710 & 0.294 & unstable & regular oscillation \\
11 & 23.4 & 1.8720 & 0.664 & 0.255 & unstable & leaves domain \\
12 & 19.9 & 1.9200 & 0.372 & 0.074 & unstable & ----- \\
13 & 34.8 & 1.9400 & 0.405 & 0.034 & unstable & ----- \\
14 & 16.8 & 1.8900 & 0.481 & 0.183 & unstable & ----- \\
15 & 19.8 & 1.8000 & 0.923 & 0.535 & stable & ----- \\
\hline
\end{tabular}
\end{center}
\end{table*}

The standard Runge-Kutta method is used to integrate differential
equations from the sonic point to the shock. In our calculations,
the adiabatic index $\gamma$ is fixed to be $4/3$ as in MTK.
Thus the lowest post-shock Mach number (corresponding to extremely
strong shocks) is: $\Msh = [(\gamma-1)/2\gamma]^{1\over 2} = 0.354$.
Tab.~\ref{case15} presents a sample of 15 shocked accretion flows,
of which cases 1-11 were exactly taken from Table 1 of MTK.
As shown in Tab.~\ref{case15}, for cases 1-11, the ranges of
the post-shock Mach number $\Msh (0.493,0.760)$ and
the minimum Mach number in the subsonic region between the sonic point
and the shock $\Mmin (0.255,0.650)$ are a little narrow
compared with the theoretical ones, $\Msh (0.354,1)$ and $\Mmin (0,1)$,
respectively.
Thus cases 12-15 are added into Tab.~\ref{case15} to make the
ranges wider, i.e. $\Msh (0.372,0.923)$ and $\Mmin (0.034,0.650)$.
Our linear results for $m=1$ and MTK simulation results 
are listed on the sixth and seventh columns of Tab.~\ref{case15},
respectively.

Fig.~\ref{figle} shows the domain $(l,e)$ of angular momentum and
Bernoulli constant for which an adiabatic flow, subsonic far from the
accretor, may be accreted onto a black-hole through a stationary shock.
Each point between the two
solid lines represents an inner shock (post-shock acceleration) and an
outer shock (post-shock deceleration).
Since the inner shock was already found to be unstable to axisymmetric
perturbations, we will concentrate on the stability of the outer
shock against non-axisymmetric perturbations.

\subsection{Linear results compared with MTK simulation results}

\begin{figure}
\centering
\includegraphics[width=8cm]{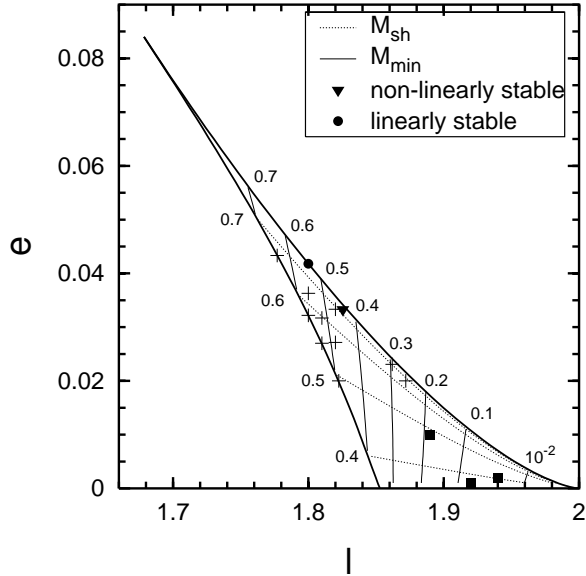}
\caption{The two thick solid lines are the threshold for shock-included
solutions. The dotted lines measure the shock strength by the value of
$\Msh$ indicated on the left. The thin solid lines correspond to
the value of the minimum Mach number $\Mmin$ indicated on
the right. The 10 crosses represent cases 1-8 and 10-11, which are
unstable in MTK simulations. The filled triangle represents case 9, which
is stable in MTK simulations. The 3 filled squares represent cases 12-14,
which together with cases 1-11 are linearly unstable.
The filled circle represents case 15, which is linearly stable.}
\label{figle}
\end{figure}

As shown in Tab.~\ref{case15}, our linear numerical calculations
find that cases 1-14 are unstable, and case 15 is stable,
whereas the non-linear simulations of MTK found that
cases 1-11 are unstable except for case 9.
The agreement of linear and non-linear results can be understood
as follows. In their simulations, $\Msh$ of the stable flow (case 9)
is the largest one (among cases 1-11).
Similarly, in our linear calculations,
$\Msh$ of the stable flow (case 15) is also the largest one
(among cases 1-15). Thus both linear and non-linear results indicate
that the shock is generally unstable to non-axisymmetric perturbations
except for $\Msh \to 1$, i.e. very weak shock.
We therefore argue that there is no essential difference between
linear and non-linear results.
This general instability is even clear from Fig.~\ref{figle}, which shows
that both the linearly stable shock (case 15, filled circle) and the
non-linearly stable shock (case 9, filled triangle) locate very close to
the right border, whereas the other unstable shocks stand everywhere
except for a very narrow region close to the right $\Msh \sim 1$ border. 

\begin{figure}
\centering
\includegraphics[width=8cm]{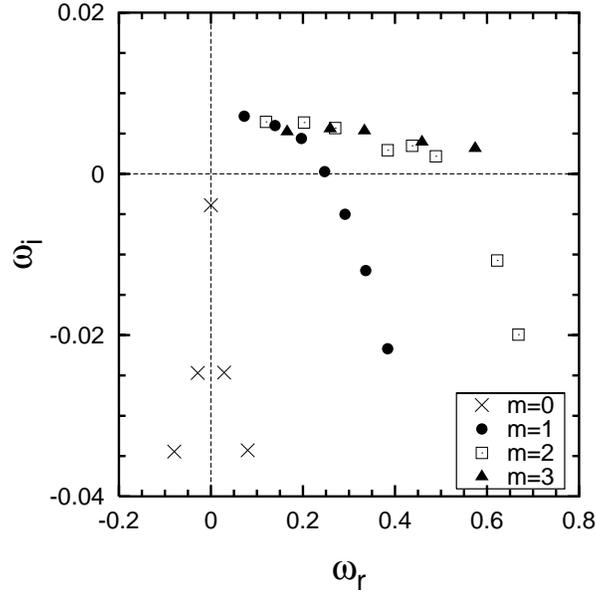}
\caption{Case 5, eigenspectrum for $0 \le m \le 3$.}
\label{figm}
\end{figure}

As a typical example, the eigenspectrum of case 5
is shown in Fig.~\ref{figm} for
perturbations $0 \le m \le 3$. The shock is linearly stable to
axisymmetric perturbations ($m=0$), which coincides with the criterion
``post-shock acceleration causes instability".  Despite the
post-shock deceleration, however, the shock is found to be unstable to
non-axisymmetric perturbations with $m=1,2,3$.
The MTK simulations found that, for case 5, the perturbed axisymmetric
shock will finally change into an $m=1$ deformed shock. Such a non-linear
evolution can be well explained by our numerically obtained eigenmodes.
As shown in Fig.~\ref{figm}, the fastest growth rate correponds to
$m=1$, $\om = 0.0725 + 0.00715i$, which indicates that the $m=1$
perturbations should dominate over the others.

\begin{figure}
\centering
\includegraphics[width=8cm]{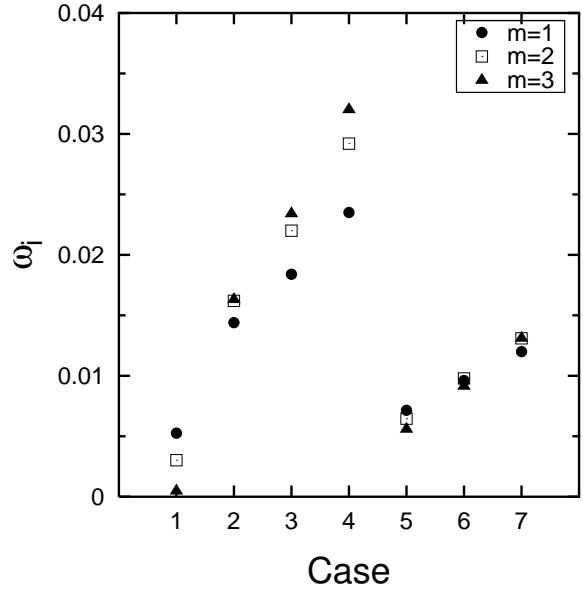}
\caption{Cases 1-7, growth rates for $1 \le m \le 3$.}
\label{figgrow}
\end{figure}

To understand the different behaviours of perturbed shocks in MTK
simulations, we focus on cases 1-7 since there exists continuous change
from 1 to 4 and from 5 to 7, respectively.
Fig.~\ref{figgrow} shows that the fastest growth rate correponds to
$m=1$ for cases 1 and 5. For the other five cases, however,
the fastest growth rate does not correpond to $m=1$.
In other words, among cases 1-7, $m=1$ perturbations dominate for
cases 1 and 5.
As shown in Tab.~\ref{case15}, MTK simulations found that the
perturbed shock will change into an $m=1$ deformed shock 
for cases 1,2 and 5.
Comparing the linear results with simulations, we therefore conclude that
the non-linear behaviours are determined by the fastest growth rate.
The $m=1$ deformed shock, i.e. ``regular oscillation", should come
into being when the fastest growth rate correponds to $m=1$.
On the contrary, other types such as ``beating" should appear when
the fastest growth rate does not correpond to $m=1$.

The only inconsistent solution is case 2. As shown in Tab.~\ref{case15}
and Fig.~\ref{figgrow}, from case 1 to 4, the transition direction
of the instability is identical for linear and non-linear results,
i.e., from $m=1$ dominance to $m\neq 1$ dominance.
The transition point locates between case 1 and 2 in our linear
calculations, however, between case 2 and 3 in MTK simulations.
As mentioned in MTK, the average distance of the final deformed shock
will be a little larger than before. Thus the outmoving of
the shock may account for the above quantitative difference.

\subsection{Instability mechanism}

\begin{figure}
\centering
\includegraphics[width=8cm]{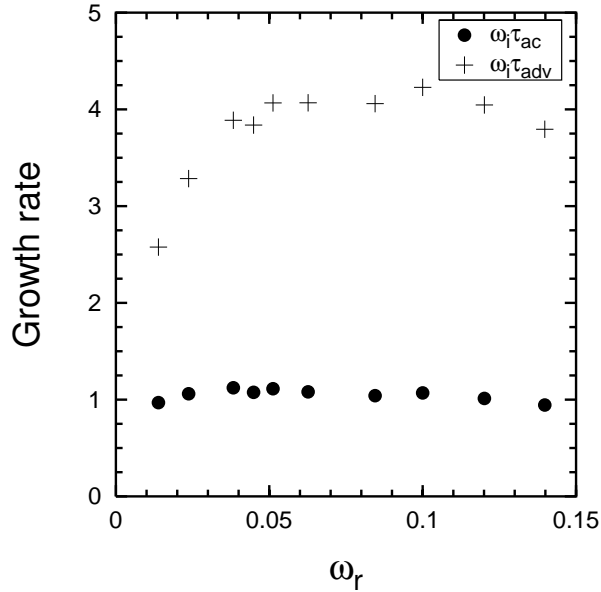}
\caption{Case 12, growth rates compared to $\Tac$ and $\Tadv$.}
\label{figwia}
\end{figure}

In the isothermal work of Paper I, the mechanism of the instability
was already found to be based on the cycle of acoustic waves between
the corotation radius and the shock.
Such a mechanism can be confirmed by the new evidence of
Fig.~\ref{figwia}, which shows 10 unstable eigenmodes for $m=1$ of
case 12. We choose case 12 because the shock is far away from
the sonic point and $\Mmin$ is low enough, thus the flow has a number of
unstable eigenmodes. 
Different from the rough timescales $\Tac$ and $\Tadv$ in Paper I,
here we numerically calculate the exact values of the time of the
purely acoustic cycle $\Tac$ and the advective-acoustic cycle $\Tadv$:
\beq
&& \Tac = \int_{\rco}^{\rsh}
\left ( \frac{1}{\cs + |\vr|} +  \frac{1}{\cs - |\vr|} \right ) \d r \ ,
\label{Tac} \\
&& \Tadv = \int_{\rco}^{\rsh}
\left ( \frac{1}{|\vr|} +  \frac{1}{\cs - |\vr|} \right ) \d r \ ,
\label{Tadv}
\eeq
where the corotation radius $\rco$ of the perturbation is defined by
$\om = m \Omega_0$, with $\Omega_0 \equiv \Omega(\rco)$:
\beq
&& \rco \equiv \left ( \frac{lm}{\om_r} \right ) ^{1\over 2} \ .
\label{rco}
\eeq
Fig.~\ref{figwia} shows that the growth time is always around
$\Tac$, i.e. $\om _i \Tac \approx 1$, but can be much shorter than
$\Tadv$, i.e. $\om _i \Tadv \ga 1$, which strongly indicates that the
mechanism is based on the purely acoustic cycle, not the
advective-acoustic cycle. However, such an evidence was not
noticed in Paper I.

\begin{figure}
\centering
\includegraphics[width=8cm]{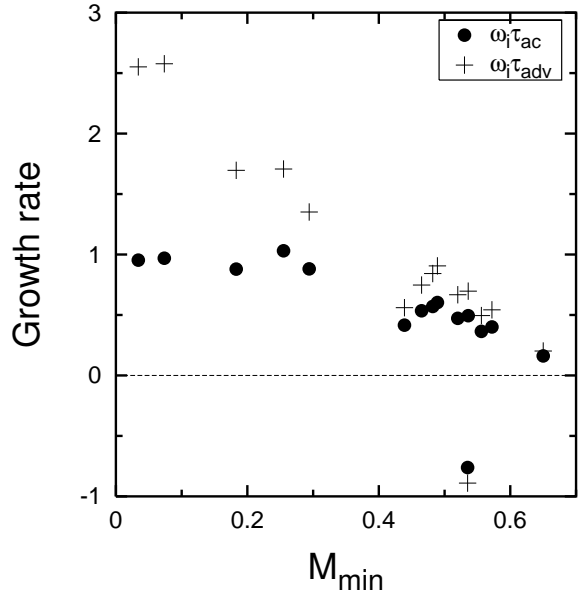}
\caption{Cases 1-15, growth rates of the most unstable mode compared
to $\Tac$ and $\Tadv$.}
\label{figwib}
\end{figure}

To have a global view, for $m=1$, the most unstable modes of all
the 15 flows are included in Fig.~\ref{figwib}, which shows that for
low $\Mmin$, i.e. $\Tac \ll \Tadv$, the most unstable mode well
matches $\om _i \Tac \approx 1$. This result again confirms the above
mentioned mechanism in adiabatic flows. The value of $\om _i \Tac$,
however, evidently decreases as $\Mmin \to 1$, which implies that
the instability is suppressed by advection.
Different from Fig.~5 of Paper I, in Fig.~\ref{figwib} here we choose
$\Mmin$ as abscissa instead of $\Msh$ because an adiabatic flow
with $\gamma = 4/3$ can not have very low $\Msh$,
and more importantly, $\Mmin$ implies the strength of advection.

\subsection{Oscillation period}

\begin{figure}
\centering
\includegraphics[width=8cm]{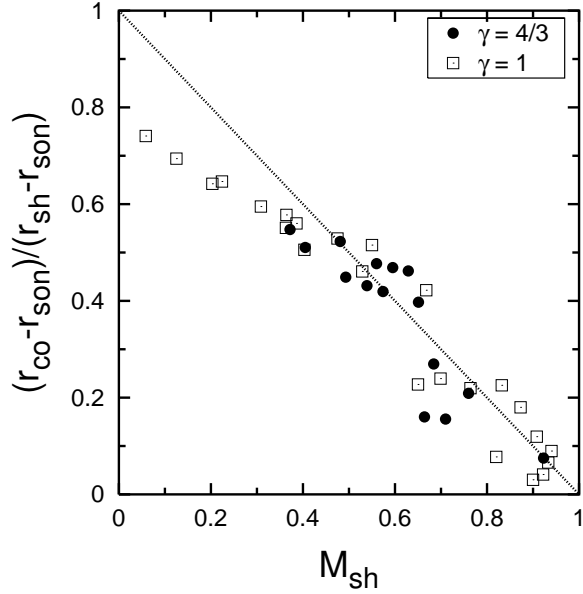}
\caption{The positions of the corotation radius of the most unstable mode
are indicated for 15 adiabatic flows (circles) and 24 isothermal flows
(squares).
The dotted line corresponds to $(\rco-\rs)/(\rsh-\rs) = 1-\Msh$.}
\label{figcoro}
\end{figure}

The growth rate is relevant to the imaginary part of the eigenfrequency
$\om_i$, whereas the oscillation period is relevant to the real part
$\om_r$ (or $\rco$).
Fig.~\ref{figcoro} includes all the 15 flows in Tab.~\ref{case15}
together with the 24 isothermal flows from Paper I, which
shows the relationship among $\rco$, $\rs$, $\rsh$ and $\Msh$.
For $\Msh$ that is not very low, Fig.~\ref{figcoro} suggests
the following good approximation:
\beq
&& \frac{\rco-\rs}{\rsh-\rs} = 1-\Msh \ .
\label{approx}
\eeq
Thus an analytical formula for calculating the oscillation period
is obtained:
\beq
&& \Pa = \frac{2\pi \rco^2}{l} 
= \frac{2\pi}{l} \left [ \rsh-(\rsh-\rs)\Msh \right ]^2 \ .
\label{Pa}
\eeq
The lowest post-shock Mach number $\Msh = [(\gamma-1)/2\gamma]^{1\over 2}$
implies that only an isothermal flow or a flow with $\gamma \to 1$ can
have extremely low value of $\Msh$. Thus Eq.~(\ref{Pa})
should work well in normal adiabatic flows with $4/3 \leq \gamma \leq 5/3$.
In addition, we define the linear period $\Pl$ for $m=1$ as follows,
\beq
&& \Pl \equiv \frac{2\pi}{\om_r^{\rm max}} \ ,
\label{Pl}
\eeq
where $\om_r^{\rm max}$ is the real part of the eigenfrequency
corresponding to the most unstable mode.

\begin{figure}
\centering
\includegraphics[width=8cm]{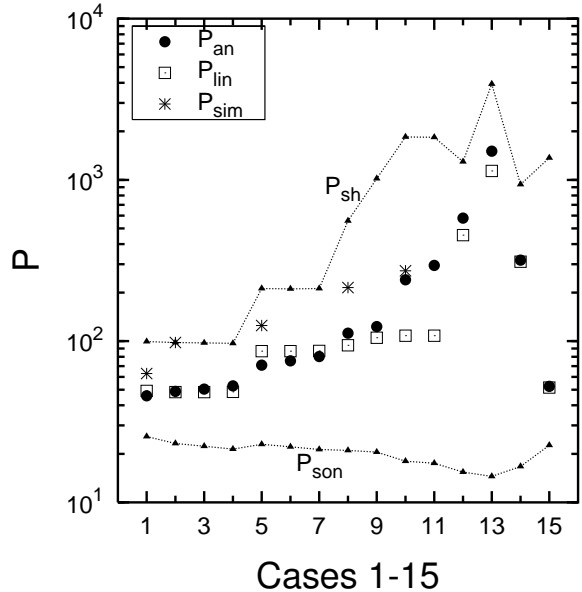}
\caption{Three types of periods for cases 1-15: the analytical $\Pa$,
the linear $\Pl$ and the MTK simulation results $\Ps$.
The upper and the lower dotted lines correspond to
$P_{\rm sh}$ and $P_{\rm son}$, respectively.}
\label{figP}
\end{figure}

The different types of periods are shown in Fig.~\ref{figP}. The lower
and the upper dotted lines correspond to the rotation period at
the sonic point $P_{\rm son} = 2\pi \rs ^2/l$ and at the shock
$P_{\rm sh} = 2\pi \rsh ^2/l$, respectively.
This figure shows that both the linear $\Pl$ and the non-linear $\Ps$
always locate between $P_{\rm son}$ and $P_{\rm sh}$ (except for case 2),
which is in good agreement with the condition that the corotation
radius should locate between the sonic point and the shock.
Fig.~\ref{figP} demonstrates that the analytical period $\Pa$
is a good approximation for the linear period $\Pl$ and
not far from the non-linear period $\Ps$.
As mentioned in MTK, the average distance of the final deformed shock
will be a little larger than before. For example, in case 5, the
final distance $\rsh$ varies between 9-11, which is larger than the
original value of 7.8. If this outmoving is taken into consideration,
$P_{\rm sh} = 2\pi \rsh ^2/l$ should therefore be even larger thus
$\Ps$ of case 2 should indeed locate between
$P_{\rm son}$ and $P_{\rm sh}$.
Futhermore, the outmoving well explains why $\Ps$ is always larger
than $\Pl$ and $\Pa$ (as shown in Fig.~\ref{figP}).

\section{Conclusions and discussion}

The main results can be summarized as follows:

\begin{enumerate}
\item
The general non-axisymmetric instability of the outer shock, which was
previously found by non-linear simulations for adiabatic flows and
linear calculations for isothermal flows, is confirmed in the present
paper by linear calculations for adiabatic flows.
\item
The simulation results of MTK are well explained by our numerically
obtained linear eigenmodes.
\item
New evidence is shown to support the argument in Paper I
for the mechanism that
the instability is based on the cycle of acoustic waves between
the corotation radius and the shock.
\item
An analytical formula is obtained to calculate the oscillation period
$P$ from the physical parameters $\rs$, $\rsh$, $l$ and $\Msh$.
\end{enumerate}

The present work is for an inviscid accretion flow with a standing
radial shock.
Realistic astrophysical accretion flows must have viscosity,
and it remains controversial whether a standing shock
can indeed form in a viscous accretion flow around a black hole.
In an accretion flow, the gravitational potential energy
is converted to kinetic and thermal energy of the accreting gas.
Based on the energy consideration, Narayan, Mahadevan \& Quataert (1998)
identified three regimes of accretion: the radiative cooling-dominated
accretion flow, the advection-dominated accretion flow (ADAF),
and the low energy generated accretion flow.
For cooling-dominated flows or ADAFs, the released gravitational
potential energy is mainly converted to thermal energy by
viscous stresses and then radiated away or advected into the central
black hole.
Except for the region very close to the
black hole ($\la 5 r_{\rm g}$), the radial motion of these two kinds of
accretion flows keeps subsonic because of the low kinetic energy.
This is the physical reason why there are no shocks in the
ADAF-thin disc solutions
(e.g. Narayan, Kato \& Honma 1997, hereafter NKH), and this result
does not rely on the mathematical technology.
In fact, Lu, Gu \& Yuan (1999) recovered the shock-free
ADAF-thin disc solutions using the Runge-Kutta method
(by adjusting the specific angular momentum $j$ accreted by the
black hole and the sonic point $\rs$ of an ADAF to match a thin disc
at the outer boundary), which are identical with those in NKH
using the relaxation method
(by assuming the value of the outer boundary $r_{\rm out}$
and the conditions at $r_{\rm out}$, and then solving the set of
euqations and calculating out $j$ and $\rs$ as eigenvalues).
In our opinion, a standing shock may form if the original flow belongs to
the third regime mentioned above,
i.e. a low energy generated accretion flow.
If the flow has very low angular momentum $l$ at large distance,
i.e. close to Bondi accretion
(in an ADAF $l$ is not very low, see Fig.2 of NKH),
the gravitational force dominates over the centrifugal force.
The flow is accelerated efficiently in the radial direction
and becomes supersonic far from the central black hole (Yuan 1999),
then a shock is likely to develop.
Such a flow is a low energy generated one because the released
gravitational potential energy is mainly converted to kinetic energy
rather than thermal energy.
NKH has also agreed that this is the situation in which
a shock can indeed be physical.
Thus we believe that flows with low angular momentum and weak
viscosity (i.e. low energy generation) will have shocks.

It is well-known that the PPI occurs with either an inner or
an outer reflecting boundary, or even more efficiently with both.
The boundary in the present paper is a standing shock.
However, a shock is not the only type of boundary which can
reflect acoustic waves. For example, the transition from a
Shakura-Sunyaev disc (Shakura \& Sunyaev 1973) to an ADAF
ought to be very sharp (NKH, Manmoto \& Kato 2000, Lu, Lin \& Gu 2004).
Such a sharp transition surface can also reflect acoustic waves,
thus the PPI may occur and result in the QPO in this system.
We therefore argue that the QPO should be a common phenomenon
in accretion flows with angular momentum.
In particular, the QPO frequencies in a 3:2 ratio in black hole X-ray
binaries (McClintock \& Remillard 2005) may be explained by this
instability for both m=2 and m=3 dominance, such as cases 2 and 7
(as shown in Fig.~\ref{figgrow}).

\section*{Acknowledgments}
We thank Thierry Foglizzo for helpful discussions.
This work is supported by the National Science Foundation of China
under Grant Nos.10233030, 10503003 and the Natural Science Foundation
of Fujian Province under Grant No.Z0514001.

\end{document}